\documentclass[aip, jcp, amsmath, amssymb, preprint]{revtex4-2}
\usepackage{graphics,dcolumn}
\usepackage{graphicx}
\usepackage{amsmath}
\usepackage{amsfonts}
\usepackage{color}
\newcommand{\ii}{\mathrm{i}}

\usepackage{mathtools}

\DeclarePairedDelimiterX\braket[2]{\langle}{\rangle}{#1 \delimsize\vert #2}
\usepackage[normalem]{ulem}
\usepackage{xcolor}
\begin{document}

%==============================================================================
\title{The relevance of degenerate states in chiral polaritonics}%Chiral polaritonics within the dipole approximation: possibilities and limitations}
%==============================================================================

\author{Carlos M. Bustamante}
 \affiliation{Max Planck Institute for the Structure and Dynamics of Matter and Center for Free-Electron Laser Science, Luruper Chaussee 149, 22761 Hamburg, Germany}
\author{Dominik Sidler}
\email{dominik.sidler@psi.ch}
 \affiliation{Max Planck Institute for the Structure and Dynamics of Matter and Center for Free-Electron Laser Science, Luruper Chaussee 149, 22761 Hamburg, Germany}
 \affiliation{Laboratory for Materials Simulations, Paul Scherrer Institut, 5232 Villigen PSI, Switzerland }
\author{Michael Ruggenthaler}
\author{Ángel Rubio}
\email{angel.rubio@mpsd.mpg.de}
 \affiliation{Max Planck Institute for the Structure and Dynamics of Matter and Center for Free-Electron Laser Science, Luruper Chaussee 149, 22761 Hamburg, Germany}
\affiliation{The Hamburg Center for Ultrafast Imaging, Luruper Chaussee 149, 22761 Hamburg, Germany}
% \email{carlos.bustamante@mpsd.mpg.de}
%\affiliation{Max Planck Institute for the Structure and Dynamics of Matter, Hamburg, Germany.}

\begin{abstract}
In this work we explore theoretically whether a parity-violating/chiral light-matter interaction is required to capture all relevant aspects of chiral polaritonics or if a parity-conserving/achiral theory is sufficient (e.g. long-wavelength/dipole approximation). This question is non-trivial to answer, since achiral theories (Hamiltonians) still possess chiral solutions. To elucidate this fundamental theoretical question, a simple GaAs quantum ring model is coupled to an effective chiral mode of a single-handedness optical cavity in dipole approximation. The bare matter GaAs quantum ring possesses a non-degenerate ground state and a doubly degenerate first excited state. The chiral or achiral nature (superpositions) of the degenerate excited states remains undetermined for an isolated matter system. However, inside our parity-conserving description of a chiral cavity, we find that the dressed eigenstates automatically (ab-initio) attain chiral character and become energetically discriminated based on the handedness of the cavity. In contrast, the non-degenerate bare matter state (ground state) does not show an energetic discrimination inside a chiral cavity within dipole approximation. Nevertheless, our results suggest that the handedness of the cavity can still be imprinted onto these states (e.g. angular momentum and chiral current densities). 
Overall, above findings highlight the relevance of degenerate states in chiral polaritonics. In particular, because recent theoretical results for linearly polarized cavities indicate the formation of a frustrated and highly-degenerate electronic ground-state under collective strong coupling conditions, which, likewise, is expected to form in chiral polaritonics and thus could be prone to chiral symmetry breaking effects.
\end{abstract}

\maketitle

\section{Introduction}

The notion of chirality, i.e., that an object can not be made identical to its mirror image upon rotations and translations, is connected to various different concepts and ideas in physics and chemistry. For instance, an \textit{achiral theory} (e.g., described by a Hamiltonian invariant under parity transformations) can have \textit{solutions with chiral character} such as in the case of the Dirac equation~\cite{greiner2000relativistic,greiner2013field}. One possible way of an achiral theory to attain chiral solutions is by spontaneous symmetry breaking, e.g., by degeneracies that allow the system to choose a solution that breaks the parity symmetry of the theory~\cite{ryder1996quantum,srednicki2007quantum}. The concept of spontaneous symmetry breaking (not necessarily leading to chirality) is ubiquitous in (theoretical) chemistry. For instance, when we localize a molecule and separate off translations and rotations, the free-space symmetry is broken. On the other hand, for \textit{chiral theories} parity symmetry is explicitly broken and hence  a discrimination is already evident on a Hamiltonian level~\cite{srednicki2007quantum}. In other words, a chiral theory treats one mirror image different to the other. In chemistry chiral theories (Hamiltonians) are typically applied to describe a chiral subsystem of interest. That is, while the whole system (universe) is described by an achiral theory such as standard non-relativistic quantum mechanics (having electrons, nuclei/ions quantized and only interacting via the Coulomb interaction in free space) this symmetry is explicitly broken within a (localized) subsystem. This can happen, for instance, when performing the Born-Oppenheimer approximation and localizing the nuclei/ions or when applying external parity-violating fields. It is on the Born-Oppenheimer level that chiral molecules (named \textit{enantiomers}) are usually investigated. Yet due to the fact that they are approximations to an achiral theory, no energy discrimination can be described without a further external mechanism (e.g., a chiral cavity). Based on this preliminary discussion, we note that there is no straightforward unified theoretical approach to describe/investigate and possibly modify chiral chemistry (i.e. the handedness of the solutions). A priori both, a chiral and an achiral theory might be possible to describe a situation of chemical interest.  %and a Besides the question whether chirality refers to the theory or its solutions, 
Things become even more fuzzy if we take into account that chiral properties also depend on the dimensionality, spacetime and the chosen representations of symmetries~\cite{ryder1996quantum,srednicki2007quantum}.

Irrespective of these theoretical subtleties, matter systems that exhibit chirality are crucial in many fields of natural science~\cite{barron2008chirality}. Specifically in chemistry, molecules that exhibit chirality (enantiomers) are of central importance. While enantiomers have similar chemical and physical properties, they react differently when interacting with other chiral objects. Hence controlling and differentiating between enantiomers is important in chemical synthesis. To do so it is common to use chiral solvents or substrates, but this practice is often expensive and, in many cases, not environmentally friendly~\cite{lorenz2014processes, martens2014purification}. An alternative is to use chiral optical fields. For instance, circularly polarized light is used for detecting chiral substances~\cite{barron2009molecular, mason2007magnetic, piepho1983group} and theoretical as well as experimental works show its potential to control chemical processes enantioselectively~\cite{hashim2019enantioselective, raucci2022chiral, fehre2019enantioselective}. More recently the idea of using optical cavities with chiral modes (chiral cavities) has been considered a promising tool for enantiomeric discrimination and manipulation~\cite{hubener2021engineering, sun2022polariton,schlawin2022cavity,voronin2022single,mauro2023, schaefer2023chiral,baranov2023toward, riso2023strong, riso2024strong}. Of specific interest is the case of \textit{collective strong coupling}~\cite{ebbesen2016hybrid,sanvitto2016road,herrera2020molecular,li2022molecular}, where a large ensemble of molecules is coupled to the modes of a cavity. 

%A chiral object lacks certain spatial symmetries, making it impossible to superimpose it on its mirror image (enantiomer) under any symmetry operation. Matter systems with this characteristic are crucial in many fields of natural science \cite{barron2008chirality}. However, since they exhibit similar or identical chemical and physical properties to their enantiomers, differentiating and manipulating them, favoring one enantiomer over another, is challenging. Generally, methodologies to achieve this are based on introducing chirality into the environment. In chemistry, it is common to use chiral solvents or substrates, but these practices are often expensive and, in many cases, not eco-friendly \cite{lorenz2014processes, martens2014purification}. An alternative approach comes from physics by using chiral optical fields. For instance, circularly polarized light is used for detecting chiral substances \cite{barron2009molecular, mason2007magnetic, piepho1983group} and theoretical and experimental works show its potential to control chemical processes enantioselectevely \cite{hashim2019enantioselective, raucci2022chiral, fehre2019enantioselective}. In recent years, the idea of using optical cavities capable of confining chiral light (chiral cavities) has been considered a promising tool for enantiomeric discrimination and manipulation, specially for molecules \cite{hubener2021engineering, sun2022polariton, schäfer2023chiral, riso2023strong}. However, this emerging field still lacks sufficient experimental and theoretical support. 

As can be anticipated from the previous theoretical ambiguities, there is no straightforward and unified approach to chose a Hamiltonian that captures all chemically relevant aspects of a molecular ensemble that couples strongly to a chiral cavity. A specific theoretical aspect that we want to investigate in this work is whether we need to use an explicit parity-violating light-matter coupling term in the Hamiltonian or whether certain chiral properties can also be captured due to solutions with parity-conserving light-matter coupling terms. In particular, choosing a parity-conserving light-matter coupling term may be computationally beneficial when aiming for collective strong-coupling simulations with large molecular ensembles. The prime example of such a computationally efficient parity-conserving coupling term, even for chiral cavity modes, is the \textit{long-wavelength/dipole light-matter coupling}~\cite{jestadt2019light}. The obvious advantage of a parity-violating light-matter coupling term for polaritonics is that it allows to energetically discriminate between different enantiomers in a chiral cavity even for non-degenerate states. If, however, the light-matter coupling term is parity conserving, i.e., the interaction with a chiral cavity transforms back onto itself under parity, then both enantiomers will have the same energy in the chiral cavity. Nevertheless, this only tells that there is no energetic discrimination for non-degenerate states, but still the wave function or other observables may have chiral character. Furthermore, since we are usually interested in the collective-coupling case, which implies highly degenerate ensemble states~\cite{sidler2024unraveling}, cavity-induced spontaneous symmetry breaking becomes plausible. This in turn can affect the chiral properties of the solutions, and might become the dominating effect for chiral polaritonics. In other words, even a parity-conserving coupling term might be enough to capture the essentials of chiral polaritonics in the collective case.

Taking a step towards this direction, in this work we consider a simple toy system and investigate to which extent we need a parity-violating light-matter coupling term to describe/control chiral properties. We do so for an achiral matter system that is coupled to a single-handed chiral  cavity field, \cite{voronin2022single,mauro2023}  within the \textit{local strong coupling} regime. That is, we do not consider a large ensemble that couples collectively to the cavity mode, but instead a single entity that couples strongly. To be specific, we consider a two-dimensional quantum ring that couples to an effective single-mode cavity. This can be taken as a model for a single quantum ring coupled strongly to a micro- or nano-cavity or a single quantum ring out of an ensemble under collective strong coupling conditions. In the later case,  it has been shown numerically that disorder can induce non-negligible local effects~\cite{sidler2020polaritonic,sidler2024unraveling} that one can mimic by computationally accessible single entity strong coupling simulations  . We will investigate how various observables behave and to which extent they differ between a parity-violating and parity-conserving light-matter coupling. In the end we will discuss to which extent these results can help to anticipate what happens for more realistic setups and in the collective-coupling case of chiral polaritonics.

\section{Theory and methodology}
\label{sec:theory}

In the following we consider the usual Pauli-Fierz description ubiquitous in polaritonics, but discard for simplicity the Stern-Gerlach term~\cite{ruggenthaler2023understanding}. This leads for a single quantized particle to
\begin{equation}\label{eq:PauliFierz}
    \hat{H}  = \frac{1}{2 m}\left(-\ii \hbar \boldsymbol{\nabla} + \frac{|e|}{c} \hat{\mathbf{A}}(\mathbf{r})\right)^2 + V_{\rm ext}(\mathbf{r}) + \sum_{\alpha} \hbar \omega_{\alpha} \left( \hat{a}_{\alpha}^\dagger \hat{a}_{\alpha} +\frac{1}{2} \right),
\end{equation}
where $m$ is the mass of the charged particle, $|e|$ its charge and the free-space field consists of $\alpha = (\boldsymbol{k},s)$ modes for the wave number $\boldsymbol{k}$ and polarization $s$ in agreement with the allowed wave numbers of the matter system, to reflect the fundamental gauge principle~\cite{ruggenthaler2023understanding}. Here the consistent vector potential operator (for three dimensions) is
\begin{equation}\label{eq:vectorpotential}
    \hat{\mathbf{A}}(\mathbf{r}) = \sum_{\mathbf{k},s}  \underbrace{\sqrt{\frac{\hbar c^2}{\epsilon_0 L^3}} \frac{1}{\sqrt{2 \omega_\mathbf{k}}} \left(\boldsymbol{\epsilon}_{\mathbf{k},s} \, \hat{a}_{\mathbf{k},s} e^{\ii \mathbf{k} \cdot \mathbf{r}} +  \boldsymbol{\epsilon}^*_{\mathbf{k},s} \,\hat{a}_{\mathbf{k},s}^\dagger e^{-\ii \mathbf{k}\cdot \mathbf{r}}\right)}_{= \hat{\mathbf{A}}_{\mathbf{k},s}(\mathbf{r})},
\end{equation}
where $L^3$ is the corresponding quantization volume, $\epsilon_0$ the free-space permittivity and $c$ the free-space speed of light. Moreover, $\hat{a}_{\mathbf{k},s}$ and $\hat{a}^\dagger_{\mathbf{k},s}$ are the creation and annihilation operators for mode $(\mathbf{k},s)$ with frequency $\omega_\mathbf{k}=c|\mathbf{k}|$, and the chiral transverse (we assume Coulomb gauge) polarization vectors obey~\cite{ruggenthaler2023understanding}
\begin{equation}
    \boldsymbol{\epsilon}_{\mathbf{k},\pm} = \boldsymbol{\epsilon}_{\mathbf{k},\mp}^{*} = -\boldsymbol{\epsilon}_{\mathbf{-k},\mp}  \quad \textrm{and} \quad \boldsymbol{\epsilon}_{\mathbf{k},\pm} \cdot \boldsymbol{\epsilon}_{\mathbf{k},\pm}^{*} = 1.
\end{equation}
Thus we have quantized the photon field in a chiral representation. The connection to the linear-polarized quantization is given by expressing
\begin{align}
\hat{a}_{\mathbf{k},\pm} = \frac{1}{\sqrt{2}}\left(\hat{a}_{\mathbf{k},1} \mp \ii \hat{a}_{\mathbf{k},2}\right),
\end{align}
with $\hat{a}_{\mathbf{k},l}$ the linearly-polarized annihilation operators, where $l \in \{1,2\}$, and accordingly for the creation operators. For the linearly-polarized polarization vectors we choose explicitly $\boldsymbol{\epsilon}_{\mathbf{k},1} = (\mathbf{e}_x \times \mathbf{k})/|\mathbf{e}_x \times \mathbf{k}|$ and $\boldsymbol{\epsilon}_{\mathbf{k},2} = [\mathbf{k}\times (\mathbf{e}_x \times \mathbf{k})]/|\mathbf{k}\times (\mathbf{e}_x \times \mathbf{k})|$ provided $\mathbf{k} \not \propto \mathbf{e}_x$, and accordingly otherwise. 
We have added an external scalar potential as $V_{\rm ext}(\mathbf{r})$ and we can add an external classical vector potential $\mathbf{A}_{\mathrm{ext}}(\mathbf{r})$ by replacing
\begin{align}
\hat{\mathbf{A}}(\mathbf{r}) \rightarrow \hat{\mathbf{A}}(\mathbf{r}) + \mathbf{A}_{\mathrm{ext}}(\mathbf{r}).
\end{align}
If we next want to investigate the effect of a chiral cavity, we can imagine that some of the free-space modes are enhanced due to a photonic environment.\cite{voronin2022single,mauro2023} That is, similar to linear polarization and the usual Fabry-P\'erot cavity~\cite{ruggenthaler2023understanding}, we assume that we have a local re-distribution of the photonic density of states to certain chiral modes, while the rest of the (quasi) continuum of states is subsumed in the observable mass of the charged particle. This procedure avoids some intrinsic gauge-inconsistencies with respect to the free-space description of matter, but several subtle points about the description of realistic cavities remain~\cite{ruggenthaler2023understanding}.

\begin{figure}
    \centering
    \includegraphics[scale=0.4]{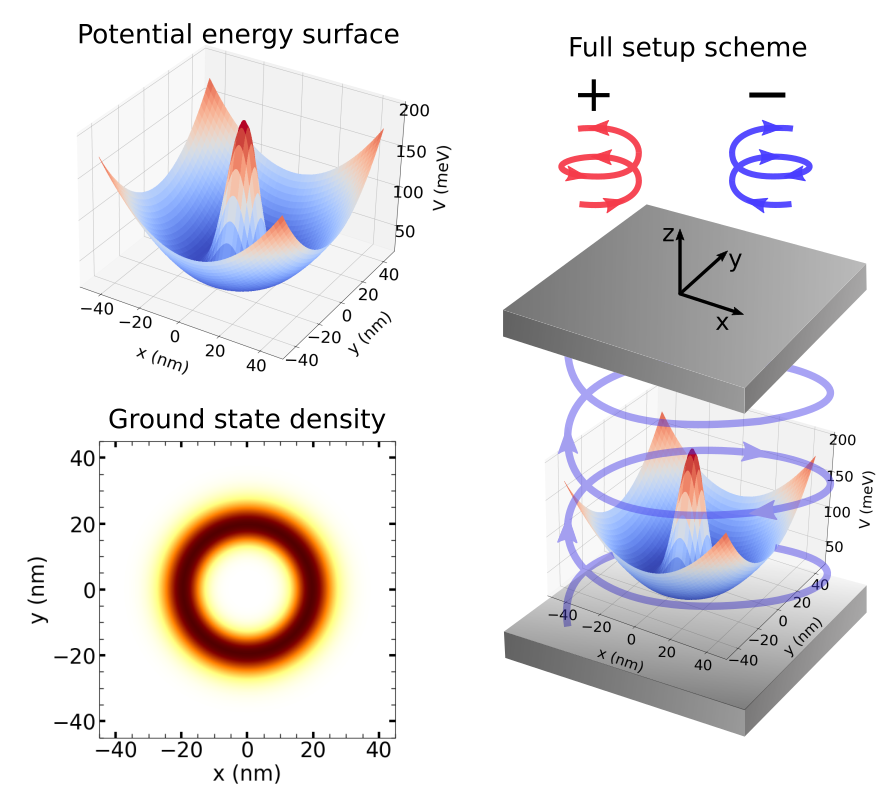}
    \caption{We consider a single effective electron that is restricted to two dimensions in an external "Mexican-hat"-like potential (top left), which gives rice to a quantum ring electronic density (bottom left). The effective electron interacts with a quantized chiral photon mode polarized along the $z$-axis (right). The different handedness of the cavity is represented by their corresponding spiral orientation.}
    \label{fig:hat_pot}
\end{figure}

Irrespective of these subtle points, we will now choose a special case of this general procedure by first restricting the matter subsystem to two effective dimensions, i.e., the electron is strongly confined within one dimension (in our case in $z$ direction) and only free to move in the two perpendicular ones (in our case in the $x-y$ plane). We will choose a model of a GaAs quantum ring, which is modelled as a "Mexican-hat" potential given by
\begin{equation}
    V_{\rm ext}(\mathbf{r}) = \frac{1}{2}m\nu^2\mathbf{r}^2 + V_0 e^{-\mathbf{r}^2/d^2}.
\end{equation}
Our potential choice indicates an achiral bare matter theory (uncoupled matter Hamiltonian), since under parity transformation one finds $\boldsymbol{\nabla}^2 = \boldsymbol{\nabla} \cdot \boldsymbol{\nabla} \rightarrow  \boldsymbol{\nabla}^2$ and $V_{\rm ext}(\mathbf{r}) \rightarrow V_{\rm ext}(-\mathbf{r})= V_{\rm ext}(\mathbf{r})$.
Here $\mathbf{r} = (x,y,0)$ and $\hbar \nu=10$ meV, $V_0=200$ meV, $d=10$ nm, as well as $m=0.067 m_e$, with $m_e$ being the mass of one electron. These parameters give the expected energy and length scales of the GaAs quantum ring~\cite{rasanen2007optimal}. For the cavity structure we envision a setup where the effectively enhanced modes due to the cavity are proportional to $\mathbf{k}=(0,0,k)$ such that the polarization vectors become 
\begin{equation}
    \boldsymbol{\epsilon}_{\mathbf{k},\pm} = \sqrt{\frac{1}{2}} (\boldsymbol{\epsilon}_x \pm \ii \boldsymbol{\epsilon}_y)
\end{equation}
for $k>0$. In Fig.~\ref{fig:hat_pot} the bare matter potential energy surface is depicted, the corresponding electronic ground state density of the quantum ring as well as a scheme of the total setup including the cavity.

After having specified the setup, we can now investigate the behavior of the light-matter coupling term under parity transformation. %i.e., $\mathbf{r} \rightarrow - \mathbf{r}$. We have $-\ii \hbar \boldsymbol{\nabla} \rightarrow \ii \hbar \boldsymbol{\nabla}$ and in our case $V(\mathbf{r}) \rightarrow V(-\mathbf{r})= V(\mathbf{r})$.
In general, the quantized vector potential transforms under parity as $\hat{\mathbf{A}}(\mathbf{r}) \rightarrow - \hat{\mathbf{A}}(-\mathbf{r})$. For an achiral field (e.g. free space) the light-matter coupling in the Hamiltonian operator must be parity-conserving. Therefore, the corresponding vector potential operator must obey $\hat{\mathbf{A}}(\mathbf{r}) \rightarrow - \hat{\mathbf{A}}(\mathbf{r})$, in order to cancel the minus from the parity-transformed momentum operator, i.e., $-\ii \hbar \boldsymbol{\nabla} \rightarrow \ii \hbar \boldsymbol{\nabla}$. This is fulfilled for Eq.~\eqref{eq:vectorpotential} as long as we sum over all wave numbers $\mathbf{k}$ (up to a certain cut-off $|\mathbf{k}|\leq \Lambda$) and polarizations $s$. Alternatively, specifically if one wants to consider only a few modes or a restricted dimensional setting (as in our simulations), one needs to include for each mode $\mathbf{k}$ also $-\mathbf{k}$ and sum over all $s$, to preserve the achiral nature of the light-matter interaction. Thus to have an explicitly chiral field present, one needs to break the symmetry between $\mathbf{k}$ and $-\mathbf{k}$ or between $+$ and $-$ in the chiral mode expansion. Alternatively, one can also introduce a symmetry-breaking external field. For instance, one can include an external vector potential of the form %\charly{(I include $c$ here to get $B_0$ in units of Tesla)}
\begin{equation}\label{eq:magenticfield}
    \mathbf{A}_{\mathrm{ext}}(\mathbf{r}) = -\frac{c B_0}{2}y\mathbf{e}_x + \frac{c B_0}{2}x\mathbf{e}_y,
\end{equation}
to break the parity of the total light-matter coupling term. Since the external field transforms as a pseudovector, i.e., $\mathbf{A}_{\mathrm{ext}}(\mathbf{r}) \rightarrow \mathbf{A}_{\mathrm{ext}}(\mathbf{r})$, the resulting light-matter coupling contains a parity-violating term. Let us note at this point that the minimal-coupling operator $(-\ii \hbar \boldsymbol{\nabla} + \frac{|e|}{c^2}\hat{\mathbf{A}}(\mathbf{r}))^2$ is defined more precisely via $(-\ii \hbar \boldsymbol{\nabla} + \frac{|e|}{c^2}\hat{\mathbf{A}}(\mathbf{r}))\cdot(-\ii \hbar \boldsymbol{\nabla} + \frac{|e|}{c^2}\hat{\mathbf{A}}(\mathbf{r}))$. Therefore the parity transformation is acting on the individual vector-valued operators of the Euclidean inner product. And only if $\hat{\mathbf{A}}(\mathbf{r})$ transforms as $-\ii \hbar \boldsymbol{\nabla}$, i.e., odd under parity, the full expression conserves parity. This behavior can be seen from the fact that the term $\hat{\mathbf{A}}({\mathbf{r}})\cdot (-\ii \hbar \boldsymbol{\nabla})$ (in Coulomb gauge) probes the phase of the wave function $\Psi(\mathbf{r})=\sqrt{\rho(\mathbf{r})} \,\mathrm{exp}(\ii \chi(\mathbf{r}))$ and vanishes if the wave function is purely real-valued. Since the difference between an even or odd complex-valued wave function is due to a phase-jump through zero, in both cases $\boldsymbol{\nabla}\chi(\mathbf{r})$ is odd away from zero. Similar to the scalar case, where the expectation value of every odd operator gives zero for pure even/odd states (because it maps even $\leftrightarrow$ odd), so does a chiral vector field.

Let us next consider a simplified description of a chiral cavity. Here we focus on the long-wavelength/dipole approximation. That is, we assume that the wavelength of the cavity-enhanced modes is small compared to the spatial extent of the matter system such that we can approximate $e^{\ii \mathbf{k}\cdot \mathbf{r}} \approx 1$. This approximation introduces extra symmetries. On the one hand, for each mode in Eq.~\eqref{eq:vectorpotential} we have now
\begin{align}
\hat{\mathbf{A}}_{\mathbf{k},s} \rightarrow - \hat{\mathbf{A}}_{\mathbf{k},s},
\end{align}
because only the polarization vectors are transformed upon parity. On the other hand, we can find combinations of quantized modes such that they transform in a chiral as well as an achiral manner, e.g.,
\begin{align}\label{eq:chiraldipole}
\left(\hat{\mathbf{A}}_{\mathbf{k},+} + \hat{\mathbf{A}}_{-\mathbf{k},-}\right) \rightarrow  - \left(\hat{\mathbf{A}}_{\mathbf{k},+} + \hat{\mathbf{A}}_{-\mathbf{k},-} \right) = \left(\hat{\mathbf{A}}_{-\mathbf{k},-} + \hat{\mathbf{A}}_{\mathbf{k},+}\right),
\end{align}
due to using $\boldsymbol{\epsilon}_{\mathbf{k},\pm} = - \boldsymbol{\epsilon}_{-\mathbf{k},\mp}$ and just relabeling $(\mathbf{k},+) \leftrightarrow (-\mathbf{k},-)$. This ambiguity due to the missing spatial dependence is not unexpected. For the classical case, all spatially constant vector fields are achiral, since they are merely a vector. It is the quantization of the modes that allows a dipole-approximated field to be \textit{interpreted} as either chiral or achiral. Therefore the long-wavelength/dipole approximation is \textit{not explicitly parity-violating} still the handedness of the cavity remains imprinted. Consequently, we will  subsequently deal with an achiral theoretical description of a quantum ring strongly coupled to a chiral cavity.

Moreover, note that if we considered a chiral field consisting of two modes in the long-wavelength approximation as in Eq.~\eqref{eq:chiraldipole}, we can rewrite this into a single coupled mode of the form
\begin{align}
\hat{b}_{\mathbf{k},+} = \frac{1}{\sqrt{2}}\left(\hat{a}_{\mathbf{k},+} - \hat{a}_{(-\mathbf{k}),-} \right)
\end{align}
and an uncoupled mode of the form $\hat{b}_{\mathbf{k},-} = \frac{1}{\sqrt{2}}\left(\hat{a}_{\mathbf{k},+} + \hat{a}_{(-\mathbf{k}),-} \right)$ and accordingly for the creation operators. Thus, without loss of generality, we can merely consider a single circularly-polarized mode in dipole approximation in the following.

In order to now compare an explicitly parity-violating and parity-non-violating description of our chiral-cavity setup, we in the following use a single circularly-polarized mode (with either handedness + or - as shown in Fig.~\ref{fig:hat_pot}) to mimic a single-handedness\cite{voronin2022single,mauro2023} chiral cavity. and for the explicitly parity-violating interaction we include the external magnetic field of Eq.~\eqref{eq:magenticfield}. We will compare to which extent these two descriptions of a chiral-cavity setup differ and to which extent the parity-conserving description of just a circularly-polarized mode in dipole approximation can capture/control the chiral character of the coupled wave functions and their observables.

For our calculations, unless otherwise stated, we use a frequency of of the chiral cavity mode $\hbar \omega = 1.413$ meV, which is in resonance with the first transition of the uncoupled matter system. To simulate the system, we worked in a two-dimensional real-space grid of 71×71 points, with a separation of 1.27 nm between each pair of points, and 9 photon states. The chosen grid separation provides  converged energies at sufficient computational speed. The energy and wave functions of each state were solved by numerically exact diagonalization of the full Hamiltonian operator. All the observables presented in this work where calculated by numerical integration of the corresponding exact eigenstates.

\section{Results}

\subsection{Uncoupled matter system}
\label{sec:resultsA}

\begin{figure}
    \centering
    \includegraphics[scale=0.4]{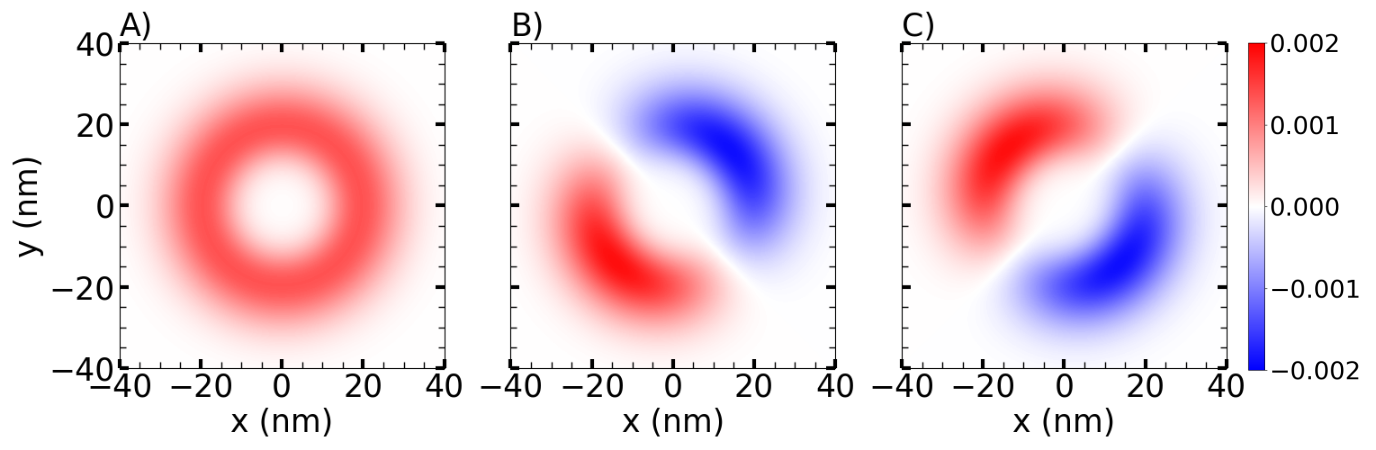}
    \caption{ Real-valued bare matter wave function of the A) ground state, B) $|M_1\rangle$ state and C) $|M_2\rangle$ state which are part of the first excited state in our uncoupled quantum-ring setup.}
    \label{fig:dens}
\end{figure}

\begin{figure}
    \centering
    \includegraphics[scale=0.4]{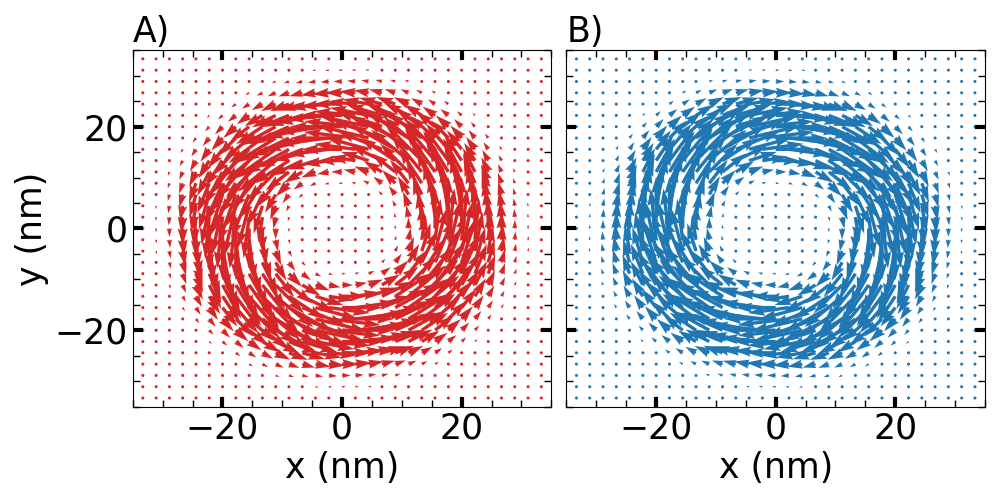}
    \caption{Excited state current densities of the uncoupled (no cavity or magnetic field) states A) $|M_+\rangle$ with $L_z = \hbar$, and B) $|M_-\rangle$ with $L_z = -\hbar$.}
    \label{fig:currents1}
\end{figure}

\begin{figure}
    \centering
    \includegraphics[scale=0.45]{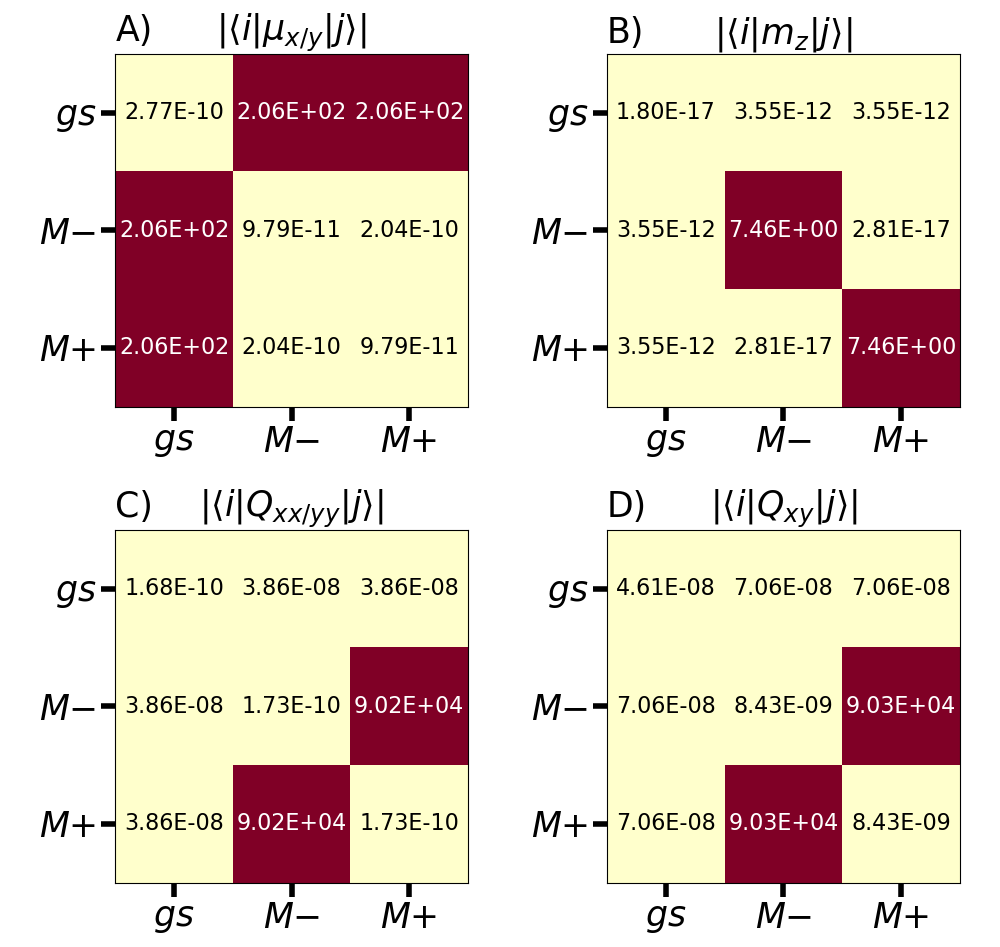}
    \caption{Magnitude of the transition terms, in atomic units for the bare system, for A) the $x$ or $y$ components of the dipole moment, B) the $z$ component of the magnetic dipole moment, C) the $xx$ or $yy$ components of the quadrupole tensor, and D) the $xy$ component of the quadrupole tensor, between the ground state (gs) and the states $M_\pm$. The main contributions are highlighted in red.}
    \label{fig:tran_mat_GS}
\end{figure}

In the absence of the cavity and magnetic field the matter Hamiltonian is symmetric under parity transformation and thus the matter-only theory is an achiral theory. Its wave-function solution shows no chiral character in the ground state. On the other hand, all the excited states of our system are doubly degenerate \cite{rasanen2007optimal}. We will restrict our focus in the following on the first two degenerate states, $|M_1\rangle$ and $|M_2\rangle$ (see Fig.~\ref{fig:dens}), which correspond to a transition of 1.413 meV for our parameter choice.
It can be seen that these states are exchangeable by a $C_4$ symmetry operation and they are odd under parity transformation. That is, these states are not identical to their mirror image but a (two-dimensional) rotation can transform them into each other. Thus these two states are not chiral in the enantiomeric (scalar and real-valued) sense. In the following the specific linear combination of these degenerate excited state
\begin{equation}
    | M_\pm \rangle = \sqrt{\frac{1}{2}} \left( | M_2 \rangle \pm \ii | M_1 \rangle \right)\label{eq:mpm}
\end{equation}
will become particularly important. The reason for this is that these states are easily accessible from the ground state in terms of angular selection rules of the electronic dipole transition operators, which impose $\Delta M=\pm 1$ given that $M=0$ in the ground state. Where $M$ corresponds to the quantum number (positive or negative), that describes the z-component of the state's angular momentum expectation value $L_z = \hbar M$.\cite{rasanen2007optimal} 
The $| M_\pm \rangle$ eigenstates are odd under parity and \textit{cannot} be transformed by a rotation around the $z$-axis into each other. Thus $|M_\pm \rangle$ are considered two-dimensional chiral excited states (solutions). The corresponding densities $\rho(\mathbf{r}) = \left| \Psi(\mathbf{r}) \right|^2 $ become cylindrically symmetric, similar to the non-degenerate ground state. However, they exhibit rotating current densities $ \left( \mathbf{J}(\mathbf{r}) = \frac{\hbar}{\ii m}\mathrm{Re}\{ \Psi^*(\mathbf{r}) \boldsymbol{\nabla}\Psi(\mathbf{r}) \} \right)$ of opposite handedness, which makes them chiral (see Fig.~\ref{fig:currents1}).
The two superpositions of opposite handedness are optically active and respond differently to circularly polarized fields. In Fig.~\ref{fig:tran_mat_GS} we can see that transitions from the ground state to the states $|M_\pm\rangle$ are dipole allowed, according to the electronic selection rules mentioned above. However, they are forbidden for magnetic dipole and electronic quadrupole transitions. In more detail, no transitions are magnetic dipole allowed, and only  transitions with $\Delta M = \pm 2$ are electrically quadrupole allowed. This is straightforward to show, due to the radial symmetry of our setup (similarly to a 2D hydrogen atom) \cite{yang1991analytic}. 

% Due to the cylindrical symmetry of our potential $V_{\rm ext}(\mathbf{r})$, the system exhibits a field-independent angular component, behaving like a rigid rotor. In this case the angular momentum is quantized according to $L_z = \hbar M$, where $M$ is an integer that can be positive, zero, or negative. The selection rules for magnetic dipole transitions and quadrupole transitions are $\Delta M = 0, \pm 1$ and $\Delta M = 0, \pm 1, \pm 2$, respectively \revMR{(This seems to contradict Fig.~4. Check wiki selection rules. there is an exception for M=0. be more precise.)}. Consequently, there is a quadrupolar transition term between $|M_-\rangle$ and $|M_+\rangle$ as can be seen in Fig.~\ref{fig:tran_mat_GS}, and no magnetic dipole transition terms, despite the fact that an energetic transition between these states is not \revMR{(really "not"? My main problem what is an "energetic" transition? Is it different to a transition?)} possible.

If we now apply a constant external magnetic field, the degeneracy of the $|M_\pm\rangle$ states is broken. Nevertheless, the angular momenta and the transition terms shown in Fig.~\ref{fig:tran_mat_GS} remain qualitatively the same (see Fig. S1). However, the explicit parity-violating interaction term can still have a physical relevant impact on the chiral states $|M_\pm\rangle$, as we will explore later in Sec. \ref{sec:resultsB}. 

\subsection{Chiral cavity without magnetic field}
\label{sec:resultsChiral}

\begin{figure}
    \centering
    \includegraphics[scale=0.4]{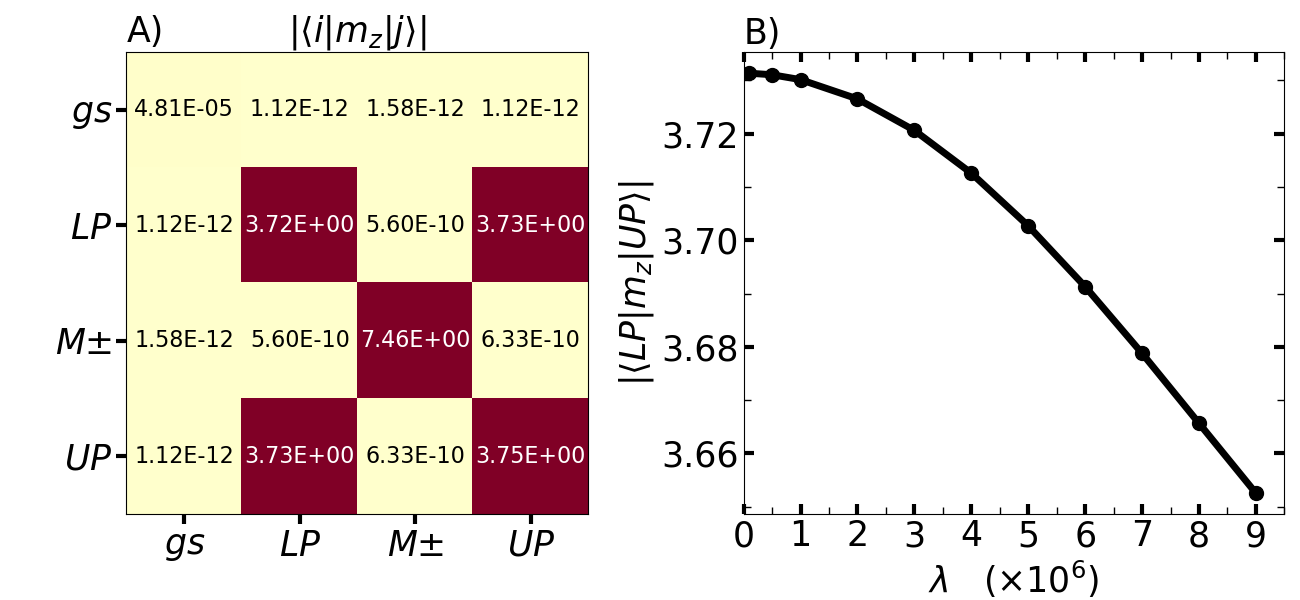}
    \caption{A) Magnetic-dipole moment ($m_z$) matrix  for a cavity-coupling of $\lambda$ = $5\times10^{-6}$ without additional external magnetic field. Relevant transitions are highlighted in red. B) Magnetic-dipole moment transition between LP and UP are plotted with respect to different coupling strengths. In both plots, the magnetic-dipole moment is expressed in atomic units. The results are independent of the cavity polarization. }
    \label{fig:mz_lamb}
\end{figure}

Next we consider how the bare matter picture changes upon coupling to a quantized effective (single) chiral cavity mode in the long-wavelength approximation. The corresponding Hamiltonian is given by,
\begin{equation}
    \hat{H}_\pm  = \frac{1}{2 m}\left(-\ii \hbar \boldsymbol{\nabla} + \frac{\lambda}{\sqrt{2\omega}} \left( \boldsymbol{\epsilon}_\pm \hat{b} + \boldsymbol{\epsilon}_\pm^* \hat{b}^{\dagger} \right)   \right)^2 + V_{\rm ext}(\mathbf{r}) + \hbar \omega \left( \hat{b}^\dagger \hat{b} +\frac{1}{2} \right),\label{eq:hamc}
\end{equation}
where $\pm$ labels the handedness of the cavity. The coupling constant between light and matter is defined as $\lambda = \sqrt{\hbar |e|^2/\epsilon_0 L^3}$, where $L^3$ corresponds to the mode volume of the cavity.
As discussed in Sec.~\ref{sec:theory}, the chosen chiral light-matter interaction will not break parity explicitly. It turns out that the cavity cannot couple equally to the two chiral eigenstates of the bare matter problem $|M_\pm\rangle$. In particular, the handedness of the cavity breaks the degeneracy of the excited states by coupling mostly with  one state and creating two polaritons. Thus a chiral cavity in dipole approximation couples selectively to the chiral superpositions  (solutions) obtained from the achiral bare matter theory (Hamiltonian). Notably, this chiral discriminating nature of the excited state couplings emerges naturally (ab-initio), whereas for the bare matter system the chiral nature only appears upon a specific choice (superposition) of the degenerate excited states. In other words, no ad hoc assumption was necessary, i.e., no a priori restriction to a  chiral basis of the bare matter system was imposed as, e.g., done in Ref. \cite{sun2022polariton} (see numerical justification below). Notice, however, the handedness of the energy spectrum   becomes only interpretable / assignable   if the handedness of a single-handed cavity is known.     
\begin{figure}
    \centering
    \includegraphics[scale=0.4]{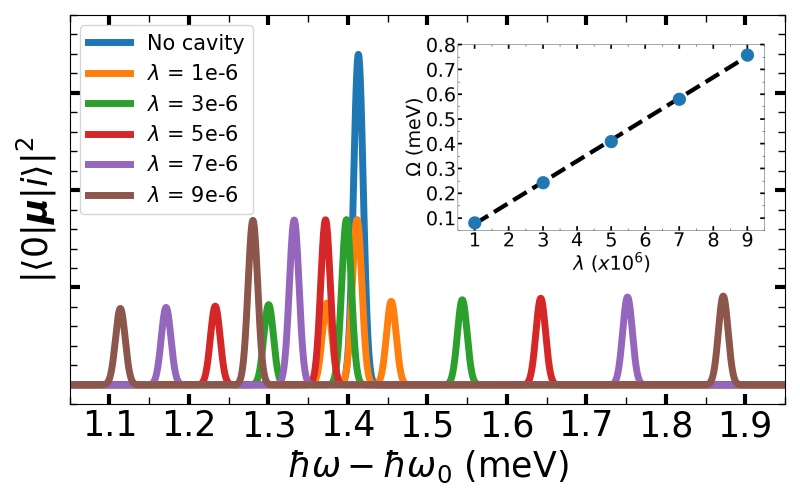}
    \caption{Linear electronic dipole absorption spectra, when tuning the chiral cavity to the first electronic excitation of the quantum ring with different coupling strengths $\lambda = \sqrt{\hbar |e|^2/\epsilon_0 L^3}$, in atomic units. The broadening of every peak has been set by a Gaussian function. The inset plot illustrates the variation of the Rabi splitting $\Omega = \hbar\omega_{UP} - \hbar\omega_{LP}$ as a function of $\lambda$.}
    \label{fig:spectrum1}
\end{figure}
In Fig.~\ref{fig:spectrum1} we present the dipole spectra for different coupling strengths ($\lambda$), by modifying $|e|^2/(\epsilon_0 L^3)$
%\revMR{(How did you fix the units? Or did you merely use atomic units in some way and recalculated everything then?)}\charly{(I worked with the expression $\lambda = \sqrt{\hbar |e|^2/\epsilon_0 L^3}$ and I used atomic units. I clarify this after this comment as well as in Fig 5. Should we use SI? If so, what are the corresponding units for $\lambda$?)} 
while keeping the frequency $\omega$ the same. In all our results, we express $\lambda$ in atomic units. Each spectrum obtained in the cavity shows three peaks: the first and third peaks correspond to the polaritonic states, and the second peak corresponds to the state $|\Psi_{m(\pm)}\rangle$ that only weakly couples %\revMR{(You wrote uncoupled. But if it shifts in energy is that correct? Is it only because the ground state is modified? Maybe explain.)} 
with the cavity  by solving Eq. \eqref{eq:hamc} numerically exactly. In more detail, investigating the projection of the bare matter chiral-superposition states $M_\pm$, defined in Eq. \eqref{eq:mpm}, reveals $|\langle M_\pm | \Psi_{m(\pm)} \rangle|^2\in [1,0.96)$ for the investigated coupling strengths $\lambda$ of a cavity described by $\hat{H}_\mp$. Due to this large overlap, we consider the middle state $|\Psi_{m(\pm)}\rangle$  as virtually unaffected by the cavity and label it, from now on, as $M_{\pm}$ (the sign is opposite to the one of the cavity). From this argument, we have justified ab-initio that a chiral cavity couples virtually only to the bare matter states that  possess the same handedness as the cavity. The resulting Rabi splitting obeys the usual linear dependency with respect to the coupling parameter $\lambda$. However, the separation of the peaks is not symmetric with respect to the cavity frequency. Similarly the peak assigned to the weakly coupled state is red shifted as we increase the coupling strength. These changes are attributed to the ``detuned'' interaction between states and the back-action of the matter system onto the cavity.\cite{sidler2024unraveling,schnappinger2023cavity,fiechter2024understanding,horak2024analytic} We will come back to this in the next section.
\\
The energy spectrum obtained by coupling with the cavity in dipole approximation does not depend on its handedness, as anticipated from our introductory discussion. However, the current density and other observables do. The later aspect is reflected in the changes of $L_z$, which we calculate according to its canonical definition
\begin{equation}
\label{eq:Lz_def}
    \hat{\mathbf{L}} = \mathbf{r} \times \left( -\ii \hbar \boldsymbol{\nabla} + \frac{|e|}{c} \hat{\mathbf{A}} \right)
\end{equation}
\begin{figure}
    \centering
    \includegraphics[scale=0.5]{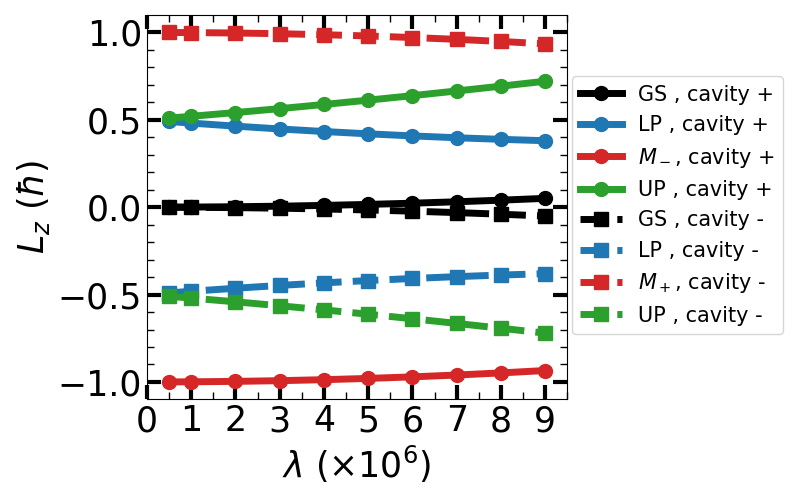}
    \caption{Angular momentum of the ground-state and the three first excited states, in the cavity (+) and (-).}
    \label{fig:Lz_lamb}
\end{figure}
In Fig.~\ref{fig:Lz_lamb} we can see the changes in $L_z$ of the ground state and the first three excited states. For lower coupling strength, the two polaritons share half of the angular momentum of the original state $|M_\pm\rangle$, while the weakly-coupled state is not considerably affected. As the coupling strength increases, $L_z$ of the upper polariton (UP) and the lower polariton (LP) separates, and the last one gets closer to the one of the ground state which moves away from zero. Thus also the ground state gets a considerable chiral character (angular currents). The value of $|L_z|$  decreases for the weakly-coupled state with respect to the coupling strength. This is because, at higher coupling strengths, the cavity imprint its opposite angular momentum. The same happens with the ground state. Furthermore, notice that the rigorous bare matter angular selection rule  are weakened by the cavity for electronic dipole transitions. Now, the transitions from the ground state to any of the first excited states involves $0 <|\Delta M|< 1$, at least within the  numerically explored parameter range of $\lambda$.
The presence of a chiral cavity in dipole approximation also enables transitions between the polaritonic states through magnetic-dipole coupling (see Fig.~\ref{fig:mz_lamb}), which is absent for the bare system. 
On the other hand, the magnitude of this transition is affected by the coupling strength of the cavity (Fig.~\ref{fig:mz_lamb}).
%For small coupling strengths, this transition follows the selection rule for rotational transitions since $\Delta M \approx 0$.
As $\lambda$ tends to 0 the transition term tends to be half of the magnetic dipole moment of the pure state (compare with Fig.~\ref{fig:tran_mat_GS}). This is because for $\lambda \neq 0$ the solution keeps being a mixture of light and matter. 
%For larger coupling strengths, the transition term persists despite $0< |\Delta M| < 1$ (see also Fig.~\ref{fig:Lz_lamb}). The violation of the uncoupled selection rule is anticipated since the state $|M_\pm\rangle$ (depending on the polarization of the cavity) is a constant component of the polaritonic states \revMR{(Do not understand, explain what you mean by this)}.
\begin{figure}
    \centering
    \includegraphics[scale=0.4]{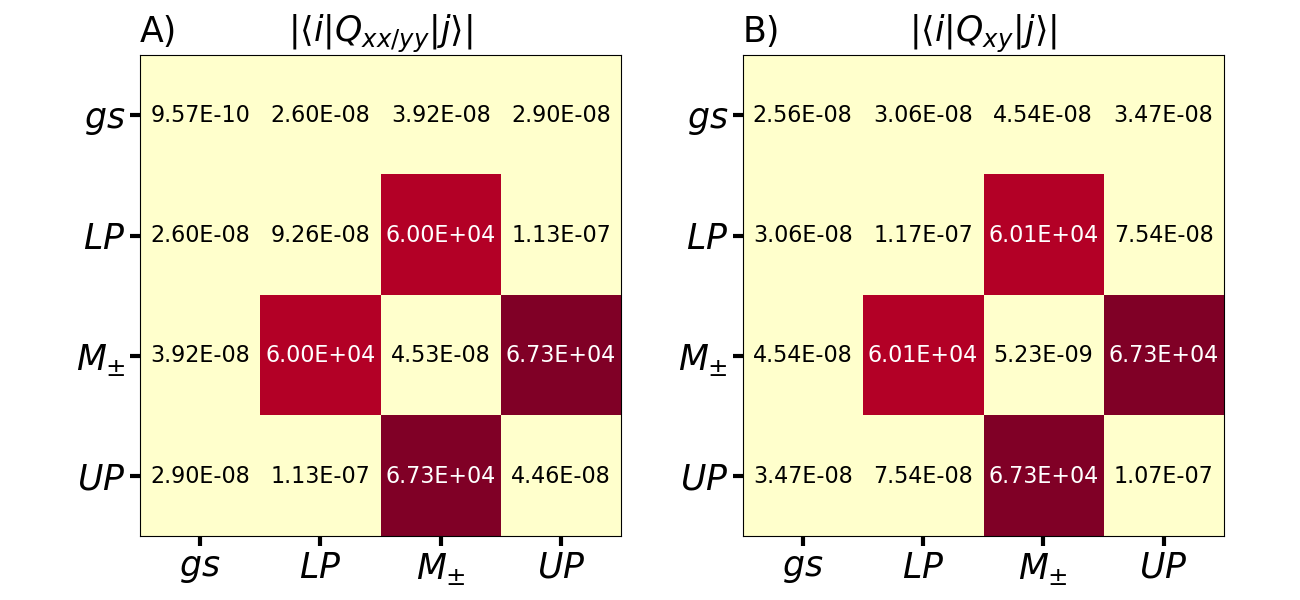}
    \caption{Matrices for the A) xx (or yy) and  B) xy components of the quadrupolar tensor, obtained for the first four states inside a cavity, using $\lambda=5\times10^{-6}$, in atomic units. No external magnetic field is applied. The shown magnitudes are independent of the cavity handedness. The main contributions are highlighted in red. The cavity-induced transitions do not follow the original selection rule $\Delta M = \pm 2$. 
    }
    \label{fig:Q_trans}
\end{figure}
%With the same argument, the existence of quadrupolar transitions between the uncoupled \revMR{(Are you sure that it is uncoupled)} state and the polaritons is expected, as shown in Fig.~\ref{fig:Q_trans}.
In Fig.~\ref{fig:Q_trans} we can see that the original quadrupole transition term between $M_+$ and $M_-$ (Fig.~\ref{fig:tran_mat_GS}) connects now the polaritonic states with $M_\pm$.
However, this suggests a violation of the bare matter selection rules due to the chiral cavity. Overall we see that already in dipole approximation various chiral properties, which are connected to violations of the uncoupled symmetries/selection rules, are present in the wave function and certain observables. 

\subsection{Chiral cavity with magnetic field}
\label{sec:resultsB}

\begin{figure}
\centering\includegraphics[scale=0.4]{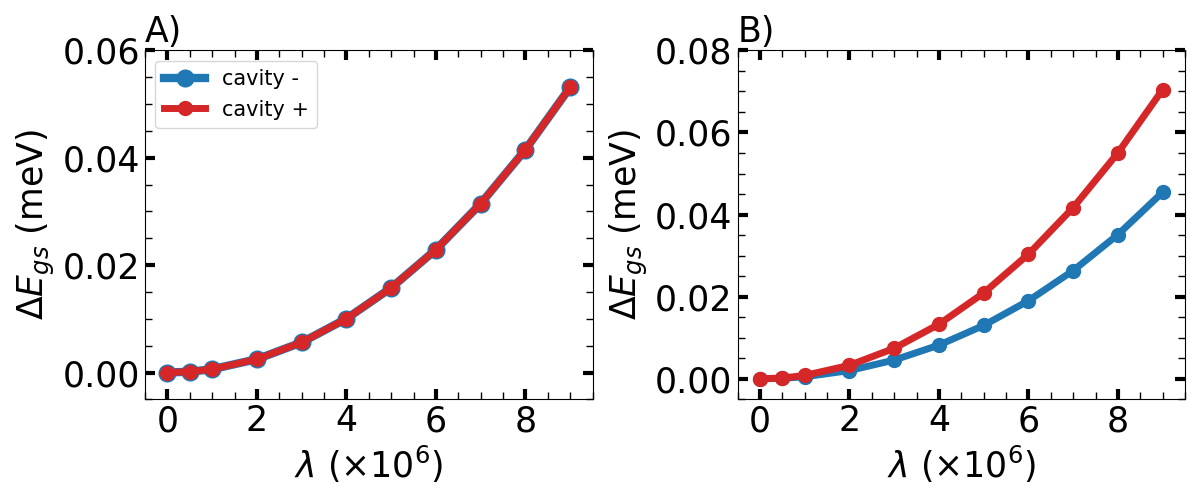}
    \caption{Comparison of the ground state energy of the system coupled with both cavities in function of $\lambda$, when A) $B_0 = 0.0$, $\hbar \omega = 1.413$ meV, and B) $B_0 = 0.235$ T, $\hbar \omega = 1.21$ meV. In both plots, $E_{gs}$ = 0.0 corresponds to the uncoupled case.}
    \label{fig:gs_discrimination}
\end{figure}

\begin{figure}
    \centering
    \includegraphics[scale=0.45]{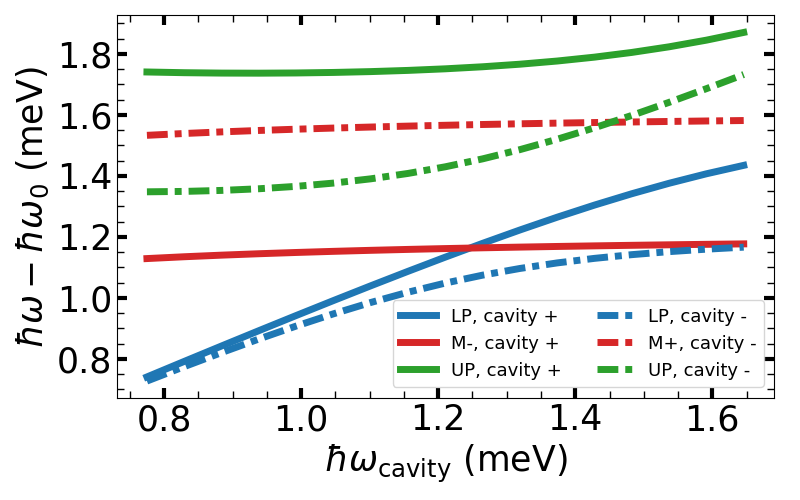}
    \caption{First three energy transitions as a function of the cavity frequency for the two different handedness, (+) or (-), where $\lambda= 5.0\times10^{-6}$, and $B_0= 0.235$ T. }
    \label{fig:avoid_cross}
\end{figure}

Next we want to see whether qualitative changes appear in the coupled light-matter wave functions and observables, once we employ an explicit parity-violating interaction term. For this purpose, we apply a static external magnetic field $B_0$ in $z$-direction. The corresponding Hamiltonian is then given
\begin{equation}
    \hat{H}_\pm  = \frac{1}{2 m}\left(-\ii \hbar \boldsymbol{\nabla} + \frac{\lambda}{\sqrt{2\omega}} \left( \boldsymbol{\epsilon}_\pm \hat{b} + \boldsymbol{\epsilon}_\pm^* \hat{b}^{\dagger} \right) - \frac{|e| B_0}{2} \left(y \mathbf{e}_x - x \mathbf{e}_y\right) \right)^2 + V_{\rm ext}(\mathbf{r}) + \hbar \omega \left( \hat{b}^\dagger \hat{b} +\frac{1}{2} \right).
\end{equation}
Explicitly adding a parity-violating interaction term, we expect that the non-degenerate bare matter ground-state is energetically discriminated by the chiral photon field. The reason is that the parity-violating part of the photon field imprints a handedness onto the non-degenerate bare-matter ground-state. It thus couples differently to the two circular polarizations of the cavity. In our simulations $B_0 = 0.235$ T was used. A comparison of the coupled ground-state energy discrimination in a parity-conserving chiral cavity with and without external magnetic fields is given in Fig. \ref{fig:gs_discrimination}.
Another interesting aspect of the externally applied magnetic field concerns its ability to reorder the excited states depending on the handedness of the cavity and the frequency of the cavity (detuning), as can be seen in Fig.~\ref{fig:avoid_cross}. In more detail,  the LP and UP can be made to even exchange ordering with the cavity-unaffected excited states $|M_\pm\rangle$.
Apart from these two aspects, the other observables (excited state discrimination, transition moments, angular momenta etc.) remain qualitatively unaffected by explicitly breaking the parity-symmetry of the light-matter interaction using a strong external magnetic field (see Figs. S2-S5 in the SI).

Importantly, at this point we need to highlight that the externally applied magnetic field is deceptively small. This is because the dimensions of our system as well as the electron's effective mass considered for our model, i.e., the magnetic flux through the quantum dot effectively determines the magnetic splitting. On molecular-sized diameters the required magnetic flux to reach significant energetic changes would be incredibly strong and thus presumably unrealistic for any practical purpose. Similarly,  beyond-dipolar contributions to the chiral cavity description (making the theory achiral)  are expected to 
 have a minuscule impact~\cite{riso2023strong}. In particular, already the much stronger dipolar coupling effects have caused substantial controversies in the polaritonic-chemistry community, whether or not they are sufficient to change chemistry locally.\cite{ruggenthaler2023understanding} We will comment on this in the next section. 

\section{Summary and conclusion}

In this work we have explored theoretically if a parity-violating light-matter coupling is required to capture all relevant aspects of chiral polaritonics or if an achiral theory is sufficient (Hamiltonian symmetric under parity transformations). This question is non-trivial to answer, since achiral theories still possess parity-violating solutions. In particular, we have demonstrated that many aspects of a chiral photon field can be captured by a non-parity violating light-matter interaction term within the long-wavelength/dipole approximation. To do so a simple GaAs quantum ring model was coupled to a non-parity violating chiral mode of a cavity. The resulting theory (coupled Hamiltonian) was achiral and thus for non-degenerate state no energetic discrimination is expected.\cite{riso2023strong}
However, things are more intricate once degeneracies are present. The bare matter GaAs quantum ring possesses a non-degenerate groundstate and a doubly degenerate first excited state. Thus the excited states allow for an achiral choice ($|M_1\rangle,|M_2\rangle$) as well as for two-dimensional chiral superpositions ($|M_+\rangle,|M_-\rangle$). In principle, those choices of excited states remain undetermined in a bare matter setup without external perturbation. However, inside our parity-conserving chiral cavity, we find that the dressed eigenstates automatically attain chiral character and become energetically discriminable, provided that the handedness of the cavity is known (fixed). In more detail, the handedness of the cavity selects and splits only one of the two degenerate chiral matter solutions into a (chiral) lower and upper polariton. The selected matter solution of opposite handedness does not split, but is slightly shifted in energy instead. 
%In contrast, the non-degenerate bare matter states (ground state) do not show an energetic discrimination inside a cavity. 
Moreover, we find that the parity-conserving chiral cavity can also imprint chiral features onto the ground state, e.g., a small angular momentum, whose handedness is determined by the cavity. However, an energetic discrimination of the non-degenerate ground state is imposed only by explicitly using a parity-violating light-matter interaction. Often an explicit energetic discrimination (parity-violating chiral theory) is achieved by going beyond the dipolar coupling of the light-matter Hamiltonian or, as in our case, by an additional external magnetic field. However, the practical issue with explicitly parity-violating light-matter contributions is their extremely small magnitude under realistic experimental conditions~\cite{riso2024strong}, which currently seems to undermine the practical feasibility of chiral polaritonic chemistry from a theoretical point of view. 

Nevertheless, our here-presented results suggest a remedy for this fundamental theoretical issue of chiral polaritonics. Instead of using numerically challenging and mathematically problematic (not every choice is consistent with the basic principles of QED~\cite{ruggenthaler2023understanding}) parity-violating light-matter theories, we propose to focus rather on simpler parity-conserving couplings in combination with highly degenerate bare matter states. We expect that such a theoretical description is much more sensitive to chiral cavity-effects, since it does not rely on beyond dipolar couplings or strong external fields. Instead, degenerate states can be re-arranged (as in the case of the excited-state manifold of our quantum-ring example) and chiral symmetries may become favorable. Our proposal is further backed by recent theoretical results for linearly polarized cavities under collective vibrational strong coupling.\cite{sidler2024unraveling,sidler2024connection} Those suggest the formation of a frustrated and highly-degenerate electronic ground state, which is prone to spontaneous symmetry breaking that impact on local molecular properties.~\cite{sidler2024unraveling,sidler2024connection} Accordingly, it seems plausible that a chiral cavity may also induce chiral symmetry breaking effects of chemical relevance. This will be the focus of future research efforts.       

\section*{Supplementary Material}
Additional simulation data is provided for the chiral cavity with an externally applied magnetic field.

\section*{Acknowledgments}
This work was made possible through the support of the RouTe
Project (13N14839), financed by the Federal Ministry of Education and Research (Bundesministerium fur Bildung und Forschung (BMBF)) and supported by the European Re-
search Council (ERC-2015-AdG694097), the Cluster of Excellence “CUI: Advanced Imaging
of Matter” of the Deutsche Forschungsgemeinschaft (DFG), EXC 2056, project ID 390715994
and the Grupos Consolidados (IT1249-19). The Flatiron Institute is a division of the Simons
Foundation. C. M. Bustamante thanks the Alexander von Humboldt-Stiftung for the financial support from Humboldt Research Fellowship. %The authors thank Dr. R. R. Riso for the fruitful discussions.

\bibliography{main}
\end{document}

%==============================================================================
\title{Supporting Information: The relevance of degenerate states in chiral polaritonics}%Chiral polaritonics within the dipole approximation: possibilities and limitations}
%==============================================================================

\author{Carlos M. Bustamante}
 \affiliation{Max Planck Institute for the Structure and Dynamics of Matter and Center for Free-Electron Laser Science, Luruper Chaussee 149, 22761 Hamburg, Germany}
\author{Dominik Sidler}
\email{dominik.sidler@psi.ch}
 \affiliation{Max Planck Institute for the Structure and Dynamics of Matter and Center for Free-Electron Laser Science, Luruper Chaussee 149, 22761 Hamburg, Germany}
 \affiliation{Laboratory for Materials Simulations, Paul Scherrer Institut, 5232 Villigen PSI, Switzerland }
\author{Michael Ruggenthaler}
\author{Ángel Rubio}
\email{angel.rubio@mpsd.mpg.de}
 \affiliation{Max Planck Institute for the Structure and Dynamics of Matter and Center for Free-Electron Laser Science, Luruper Chaussee 149, 22761 Hamburg, Germany}
\affiliation{The Hamburg Center for Ultrafast Imaging, Luruper Chaussee 149, 22761 Hamburg, Germany}
% \email{carlos.bustamante@mpsd.mpg.de}
%\affiliation{Max Planck Institute for the Structure and Dynamics of Matter, Hamburg, Germany.}

\maketitle

\begin{figure}
    \centering
    \includegraphics[scale=0.45]{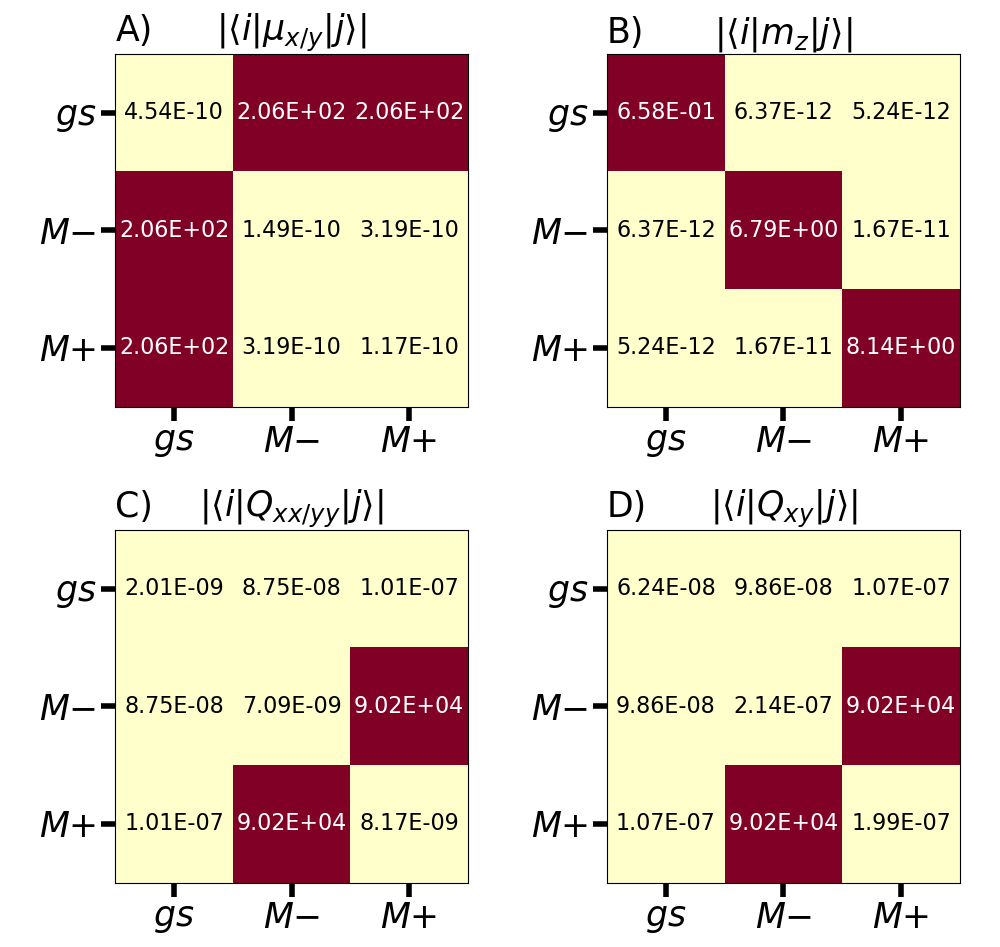}
    \caption{Magnitude of the different transition matrix elements in atomic units with an applied magnetic field ($B_0 = 0.235$ T): A)  Electronic dipole moment with respect to $x$ or $y$ components. B) The $z$ component of the magnetic dipole moment. C) The $xx$ or $yy$ components of the quadrupole tensor. D) the $xy$ component of the quadrupole tensor, between the ground state (gs) and the states $M_\pm$, which correspond to the degenerate first excited state of the uncoupled system. The mainly contributing elements of each matrix representation are highlighted in red.}
    \label{fig:tran_mat_GS}
\end{figure}

\begin{figure}
    \centering
    \includegraphics[scale=0.4]{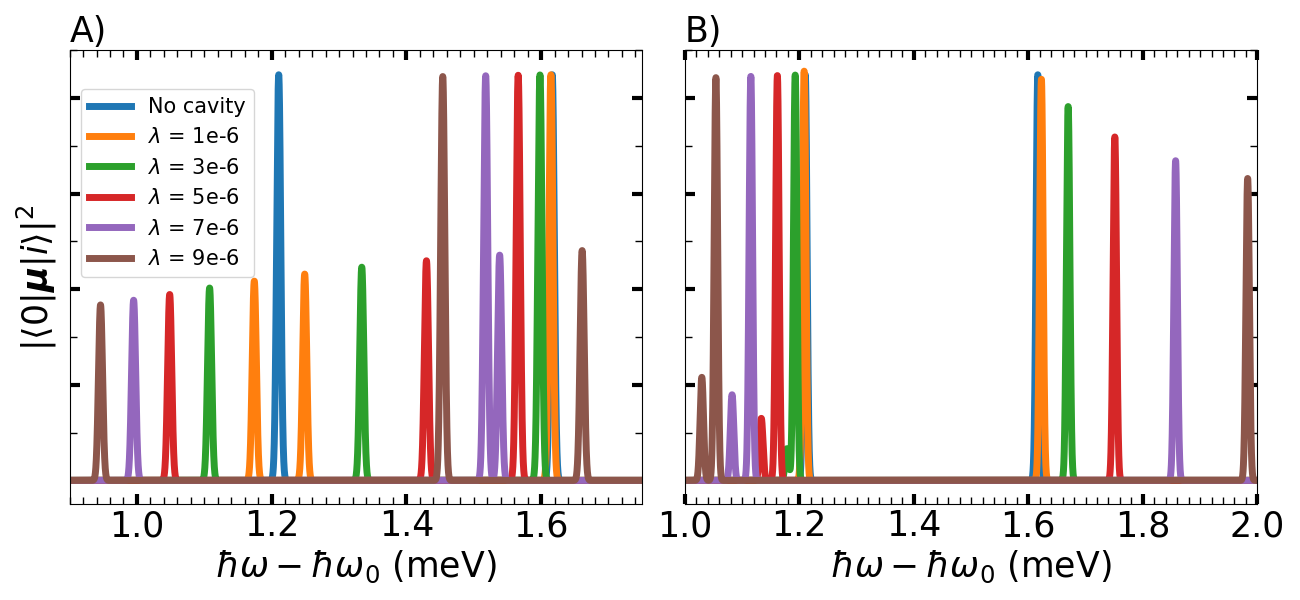}
    \caption{Linear electronic dipole absorption spectra, when tuning the A) chiral cavity (-) or B) cavity (+) to the first electronic excitation of the quantum ring. An external magnetic field is applied along the $z$-axis ($\hbar\omega = 1.21 $ meV), with different coupling strengths $\lambda$. The broadening of every peak has been set ad-hoc by a Gaussian function.}
    \label{fig:spectrum_B}
\end{figure}

\begin{figure}
    \centering
    \includegraphics[scale=0.35]{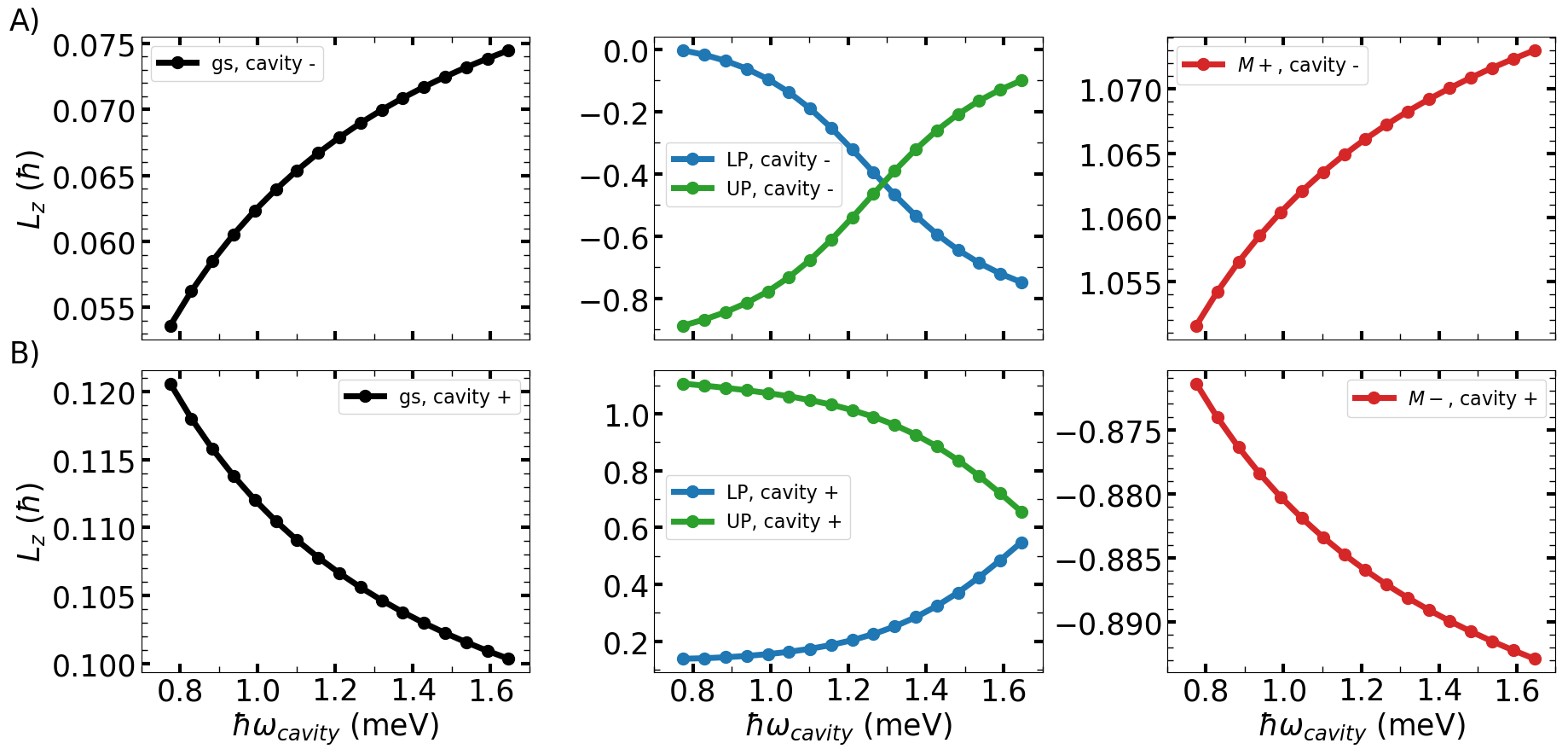}
    \caption{Angular momentum of the ground state (black), lower (blue) and upper (green) polaritons, as well as the weakly-coupled state $M_{\pm}$ (red)  obtained in a A) cavity (-) and B) cavity (+), as a function of the cavity frequency, with $\lambda= 5.0\times10^{-6}$ for $B_0 = 0.235$ T. }
    \label{fig:Lz_w0}
\end{figure}

\begin{figure}
    \centering
    \includegraphics[scale=0.35]{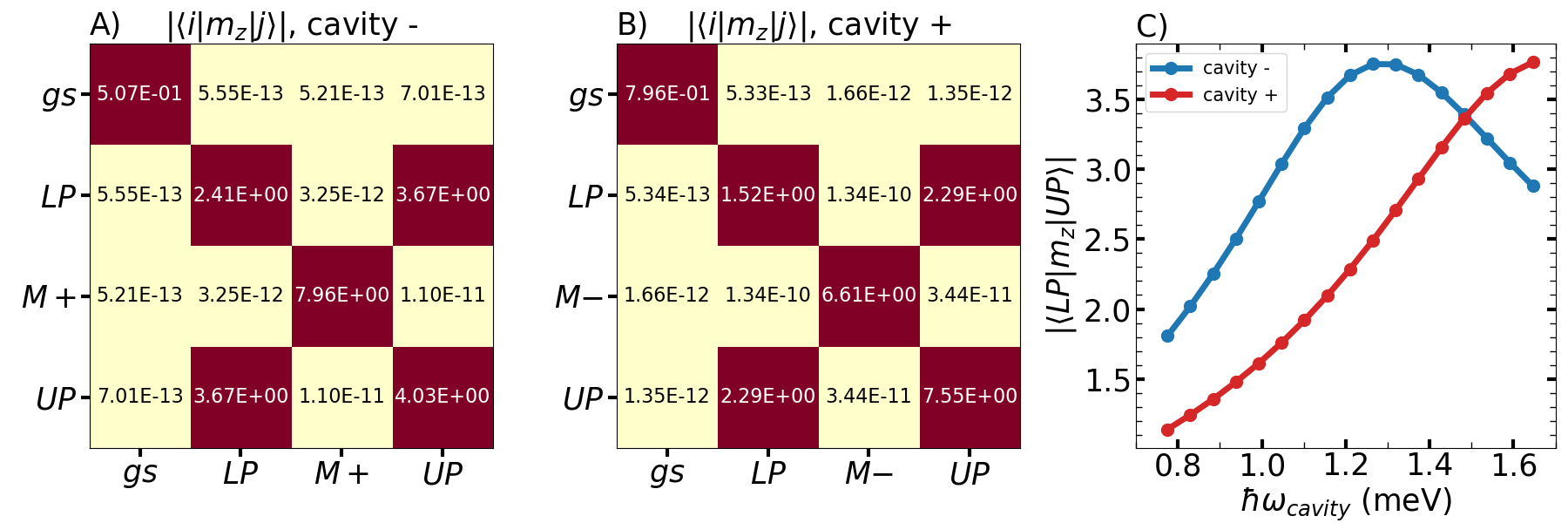}
    \caption{Magnetic-dipole transition ($m_z$) matrix obtained for a coupling strength $\lambda$ = $5\times10^{-6}$, with a cavity frequency of 1.21 meV and a constant magnetic field $B_0 = 0.235$ T. In A) the cavity corresponds to the (-) and in B) to the (+) polarization. The dominant contributions are highlighted in red. In C) we show the magnetic-dipole moment transition between LP and UP, for different cavity frequencies and polarizations, with the same $\lambda$ and magnetic field. In the three cases, the magnetic-dipole moment is expressed in atomic units. }
    \label{fig:trans_B0_w0}
\end{figure}

\begin{figure}
    \centering
    \includegraphics[scale=0.4]{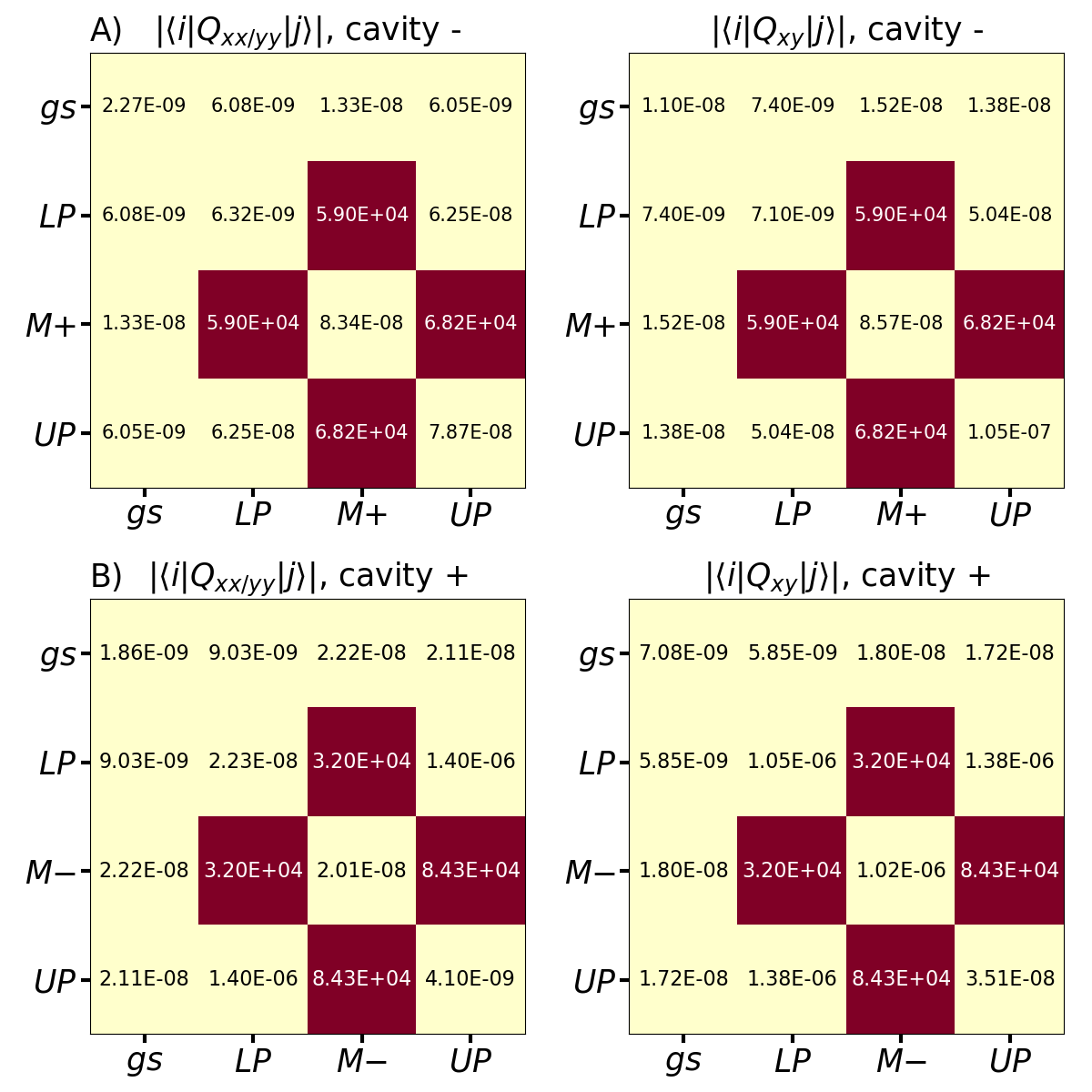}
    \caption{Matrices for the xx (or yy) and xy components of the quadropolar tensor, obtained for the first four states, when $\lambda=5\times10^{-6}$ and $B_0 = 0.235$ T, with a cavity A) (-) and B) (+) configuration. The main contributions are highlighted in red. All the magnitudes are in atomic units.}
     %\revMR{(Re-order such that we can compare with the achiral description)}
    \label{fig:Qmat_B0}
\end{figure}

%\bibliography{si}